# Orbital hybridization and electrostatic interaction in a double molecule transistor

Xiao Guo[1,2], Qing Yang[1,2,4], Wei Yu[1,3,4], Qiuhao Zhu[1,3,4], Yuwen Cai[1,3,4], Wengang Lu[1,3,4], Sheng Meng[1,2,3,4], Wenjie Liang[1,2,3,4*]

1   Beijing National center for Condensed Matter Physics, Institute of Physics, Chinese Academy of Sciences, Beijing, 100190, P. R. China
2   Songshan Lake Materials Laboratory, Dongguan, Guangdong 523808, China
3   CAS Center of Excellence in Topological Quantum Computation, University of Chinese Academy of Sciences, Beijing 100190, P.R. China.
4   School of Physical Sciences, University of Chinese Academy of Sciences, Beijing 100049, China
*Email: wjliang@iphy.ac.cn

Understanding the intermolecular interactions and utilize these interactions to effectively control the transport behavior of single molecule is the key step from single molecule device to molecular circuits[1-6]. Although many single molecule detection techniques are used to detect the molecular interaction at single-molecule level[1,4,5,7,8], probing and tuning the intermolecular interaction all by electrical approaches has not been demonstrated. In this work, we successful assemble a double molecule transistor incorporating two manganese phthalocyanine molecules, on which we probe and tune the interaction in situ by implementing electrical manipulation on molecular orbitals using gate voltage. Orbital levels of the two molecules couple to each other and couple to the universal gate differently. Electrostatic interaction is observed when single electron changing in one molecule alters the transport behavior of the other, providing the information about the dynamic process of electron sequent tunneling through a molecule. Orbital hybridization is found when two orbital levels are put into degeneracy under non-equilibrium condition, making the tunneling electrons no longer localized to a specific molecule but shared by two molecules, offering a new mechanism to control charge transfer between non-covalent molecules. Current work offer a forelook into working principles of functional electrical unit based on single molecules.

Devices constructed by single molecules not only serve as a powerful platform to investigate fundamental quantum mechanical process at molecular scale[9,10], but also exhibit wide range applications[11-16]. Thanks to the advance in nanofabrication and single molecule detection techniques, the ability to fabricate and control single molecule device has been well developed. The following task in the path towards real molecular circuit is how to combine several molecules together to form functional electrical unit, which demands the understanding interaction between molecules and its impact on the electron transport properties of molecular devices, as the intermolecular interaction will become dominate when the integrating density is high. In recent years, many attempts are performed to using single-molecule techniques to probe intrinsic molecular interaction at single-molecule level, like mechanical or SPM break junction technical[1,4,7], Scanning tunneling spectroscopy[8] and STM-induced luminescence techniques[5].

Among these studies, the intermolecular interaction can only be tuned by conformation modification induced by mechanical movement or assembling new parts into molecules[6,7]. On the demanding of molecular electrical circuits, effectively tuning the interaction in situ by manipulating molecular orbitals using electrical voltage, like what is commonly done in artificial molecule or quantum dots devices[17-22], has not been reported in real molecular devices before.

Here we assemble two individual manganese phthalocyanine (MnPc) molecules into a double molecule transistor device and probe the intermolecular interaction between the two molecules by electron transport measurement, with gate voltage used to tune the orbitals of each molecule. The whole device (Figure 1(a)) was fabricated by electromigration technique[10,23] and all the measurements were performed at 280mK in a $^3$He refrigerator (details in supplementary material).

Figure 1(b) presents the differential conductance (dI/dV) map as a function of both the bias voltage $V_{sd}$ and gate voltage $V_g$ for a representative double-molecule device. Typical coulomb blockade behavior is observed. The conductance peaks in Figure 1(b) forms two sets of coulomb diamond patterns, revealing that two individual molecules (labeled as 1 and 2) are involved in the electron transport process. The charge degeneracy points and conductance lines' slopes of two molecules are different, indicating the different local configurations of the two molecules between source and drain electrodes.

Figure 2(a) shows the detail transport map for the overlay region of two set coulomb diamond patterns(dash rectangle region in Figure 1(b)). If there is no interaction between two molecules, the transport characteristics of a double-molecule transistor would be only the combination of two individual molecules and the corresponding coulomb patterns simply overlap with each other. In Figure 2(a), however, when peak $A_1$(the negative slope line of molecule 2) extends into the sequential electron tunneling (SET) region of molecule 1, its strength decrease (become $A_2$) and another conductance peak($A_3$) appears parallelly with a bias difference $\Delta V_2$(0.74mV). For the line $B_1$(the positive slope peak of molecule 1), only an energy shift ($B_2$)is observed when it goes into the SET region of molecule 2.

These phenomena can be explained by the electrostatic interaction between two molecules, through which the potential of one molecule can be affected by the charge state of the other one. The energy diagrams of each area in Figure 2(a) are shown in Figure 2(d). When a molecule is charged by one more electron, the level of another molecule would shift a certain mount (from solid line to dash line in Figure2(d))due to the potential change. In the SET regions of molecule1, like region II, single electron tunnel into and out of molecule 1 sequentially, making it switch between two adjacent charge states. When the level of molecule 1 is occupied by the tunneling electron(left panel of II in Figure 2(d)), the level of molecule 2 shifts from the solid line to the dash line, which corresponds to peak $A_2$ in Figure 2(a). When the tunneling electron leave molecule1(right panel of II in Figure 2(d)), the level of molecule 2 goes back the solid and peak $A_3$ appears in Figure 2(a). If the level of molecule 1 is always occupied by the electron, like region I in Figure 2(a)and (d),the level is shifted all the time and only one peak $A_1$ appears. The electrostatic interaction is mutual, so the dual state of molecule 1 should also be expected when single electron tunnel through molecule2. While in the SET region of molecule2, I and II, for the reduced tunneling rate of molecule 2 towards the source electrode, the process of electron tunneling in is much faster than tunneling out, making molecule 2 is occupied by the electron all the time. So only a shift peak $B_2$ is observed.

The electrostatic interaction also makes it possible to gain deeper insight into the dynamics of electron tunneling through single molecule. In the device shown in Figure2(a), the coexisting of $A_2$ and $A_3$ in a transport diagram, which has not been shown in previous studies, provides the direct evidence to the picture of single electron sequential tunneling through a molecule when its chemical potential lies within the bias window of electrodes. The strengths of conductance peaks at the edge coulomb diamonds are proportional to the total tunneling rates of the level. The comparison of peak heights of $A_1, A_2$ and $A_3$ are shown in Figure 2(b) and the height of $A_1$ is almost the sum of the other two, indicating the total electron tunneling rate of molecule 2 keeps unchanged and doesn't depend on the charge state of molecule 1. The relative strengths of $A_2$ and $A_3$ are also indications of the average time of tunneling electron on($T_{on}$) and off($T_{off}$) molecule 1. From the strength of two peaks, we can deduce the ratio $T_{on}/T_{off}=0.34$, revealing that the electron tunnels out slightly faster than tunnels in. The total tunneling time $T_{total}(=T_{on}+T_{off})$ can be deduced by $T_{total}=|e|/I$, where I is the tunneling current of molecule 1, and I≈16.4pA in region II. Then we can estimate the average electron residence time $T_{on}$≈2.5ns and unoccupied time $T_{off}$≈7.3ns.

Further quantitative discussion about these observations can be performed based on a capacity coupled parallel double quantum dot model[24], as sketched in Figure 2(c), where single molecule working as a quantum dot,. Here, the electrostatic interaction is described by a capacitance coupling, and the chemical potentials of two molecules mutually depend on each other by the intermolecular capacitance $C_{12}$. From the voltage shift of the conductance peaks caused by charge fluctuation, we can make an estimation[20] of $C_{12}$ and $C_{12}/C_2 \sim 10^{-4}$ ($C_2$ is the total capacitance of molecule 2, details about the estimation are in supplementary material). The mutual capacitance is rather small compare to other capacitances in this molecular system.

So far, although orbital levels of the two molecules are turned respect to each other, only capacitance coupling between two molecules is observed. Electron exchange between molecules is not shown and tunneling coupling is not established. We realized this in the same device by changing local charging environment using thermal treatment (see supplementary material).

Figure 3(a) present the transport map for the same device in Figure2 after few times of thermal treatment. Molecule 1 and 2 can still be identified for the slopes of all the conductance lines are unchanged, while the charge degenerate points of both the molecules shift by a small amount. The unchanged slopes indicate the relative positions of molecule 1 and 2 with electrodes keeping the same. And the shifts of charge degenerate points probably due to the changing of local charging environment. The electrostatic interaction of two molecules is also reserved, as shown in the overlay regions at negative bias side, marked by the black dash ellipse in Figure 3(a). The shift of molecule 2's conductance peak (with negative slope) caused by the electron charging of molecule 1 is almost the same amount as that before thermal treatment (here $\Delta V_2$ =0.76mV), revealing that the mutual capacitance $C_{12}$ keep the same as well.

Detail measurement of the overlay region at positive bias side of Figure 3(a)(marked by dash rectangular) is shown in Figure 3(b). A pronounced avoid crossing is observed close the intersection point of the conductance peaks with positive slope from two molecules. This can be explained by the hybridization between two orbitals from different molecules when they both are tuned to be aligned with the Fermi level of source electrode, being degenerate. As sketched in Figure 3(c), two molecular levels from two molecules (with energy $E_1$ and $E_2$ respectively) hybridize with each other forming bonding($E_b$) and anti-bonding($E_a$) states and leaving an energy gap. The hybridized molecule orbitals extend to the entire double molecule system, forming many-body quantum states. The hybridization here is under non-equilibrium condition with single electrons tunneling sequentially, which is essentially different from the most studied equilibrium hybridization in other double quantum dot system[21,25].Upon degenerate, the electron can be exchanged not only with electrodes but also between molecules, making the tunneling electrons are no longer localized in a specific molecule but shared by two molecules.

The hybridization of two levels can be well described by quantum mechanical methods. If the hybridization strength is t, the energy of hybridized bonding state($E_b$) and anti-bonding state($E_a$) can be deduced by introducing interaction into the Hamiltonian of two-level quantum system[26,27]and expressed in terms of the original eigenvalues, $E_1$ and $E_2$ ,as follows:

$$E_b = \frac{1}{2}(E_1 + E_2) - \sqrt{\frac{1}{4}(E_1 - E_2)^2 + |t|^2},$$

$$E_a = \frac{1}{2}(E_1 + E_2) + \sqrt{\frac{1}{4}(E_1 - E_2)^2 + |t|^2}. \tag{1}$$

The plot of energy of hybridized orbitals, $E_a$ and $E_b$, which are converted from the positions of the anti-crossing conductance peaks in Figure 3(b), is shown in Figure 3(d). We use formula (1) to fit the peak position in Figure 3(d) and get t=0.083meV. The energy shift induced by single electron charging is observed again here .The dash line in Figure 3(d) indicates the energy of $E_b$ if there's no electrostatic interaction. The value of t is comparable with previous results in carbon nanotube bundles although the intermolecular capacitance here is much smaller[18,20]. This is valid as the size of single molecule is much smaller than nanotube quantum dot.

Precisely determining the configuration of the coupled molecules in the device is impossible as the ill-defined interface of electrodes created by electromigration. Form the DFT calculation, we suppose two molecule are coupled together in a partially face-to-face stacking configuration(like in Figure 1(a), details in supplementary materials). The calculated energy splitting due to hybridize and charge density difference of one particular configuration are shown in Figure 3d, whose characteristics should qualitatively apply to all the possible configurations. The charge density difference diagram shows the redistribution of electrons among two molecules upon degeneracy, confirming nonlocal molecular orbitals. The deduced distance from hybridized level splitting is also valid in our device geometry.

In the discussion above, no special molecular orbital structures are needed, so the hybridization and electrostatic interaction interactions should present in any kinds of molecules, as long as the intermolecular coupling is strong enough. Both the two interactions are observed in the first order electron tunneling process, so they can be directly used to control the charge transport of molecular circuit. Taking advantage of electrostatic interaction, single molecule can be used as charge sensor to detect the charge variation of another molecule, or as a gate to tune the potential of another molecule(like position IV and V in Figure 2(a)). In real molecular circuit future, using single atom[3] or single molecule as sensor or gate must be more flexible and effective. The hybridization provides a new mechanism to control change transfer between non covalent molecules in circuit. Tuning two levels on and off degeneracy is essentially altering the tunneling electron between local and non-local states, which can be used in molecular circuit to control direction of charge transfer or switch the circuit between on and off by adjusting the overlap of molecular orbitals using gate potential.

Molecules couple with each other in various ways, so molecular specific interactions, like dipole-dipole and exchange interaction would be expected by choosing the proper molecules. The capability of detecting and tuning the molecular interactions at single molecule level will help to gain deeper inside into the fundamental physical or chemical process at molecular scale and accelerate the pace towards real molecular circuit.

**Acknowledgements**

This research is supported by National Basic Research Program of China (2016YFA0200800), Strategic Priority Research Program of Chinese Academy of Sciences (Grant No. XDB30000000), Strategic Priority Research Program of Chinese Academy of Sciences (Grant No. XDB07030100), Sinopec Innovation Scheme (A-527), National Natural Science Foundation of China (No. 12147130), China Postdoctoral Science Foundation (No. 2021M693371).

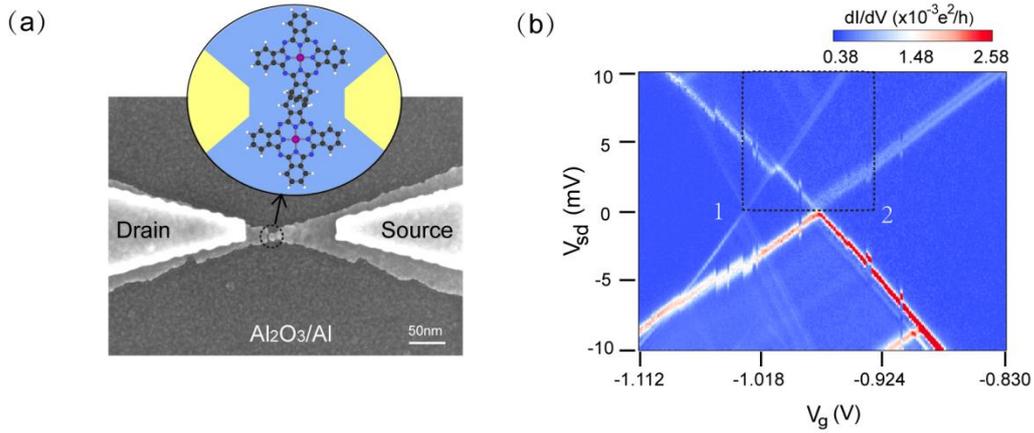

**Figure 1** (a) SEM image of a device after measurement. A gap with a separation about 1-2nm is form in gold nanowire by electromigration on the surface of $Al_2O_3/Al$ gate electrode. Insert: Structure of single Manganese(II) phthalocyanine molecule (MnPC) and schematic diagram of one possible configuration of the double MnPC molecule transistor device. (b) Color plots of differential conductance (dI/dV) as a function of bias voltage ($V_{sd}$) and gate voltage ($V_g$) for a typical double MnPc molecule transistor device at 280 mK. The conductance peaks represent the situation when molecular levels go in and leave the bias window between source and drain electrodes. Two set of coulomb diamond patterns are shown, associated with molecule 1 and molecule 2. The dash rectangle marks the overlay region of two sets Coulomb diamond patterns.

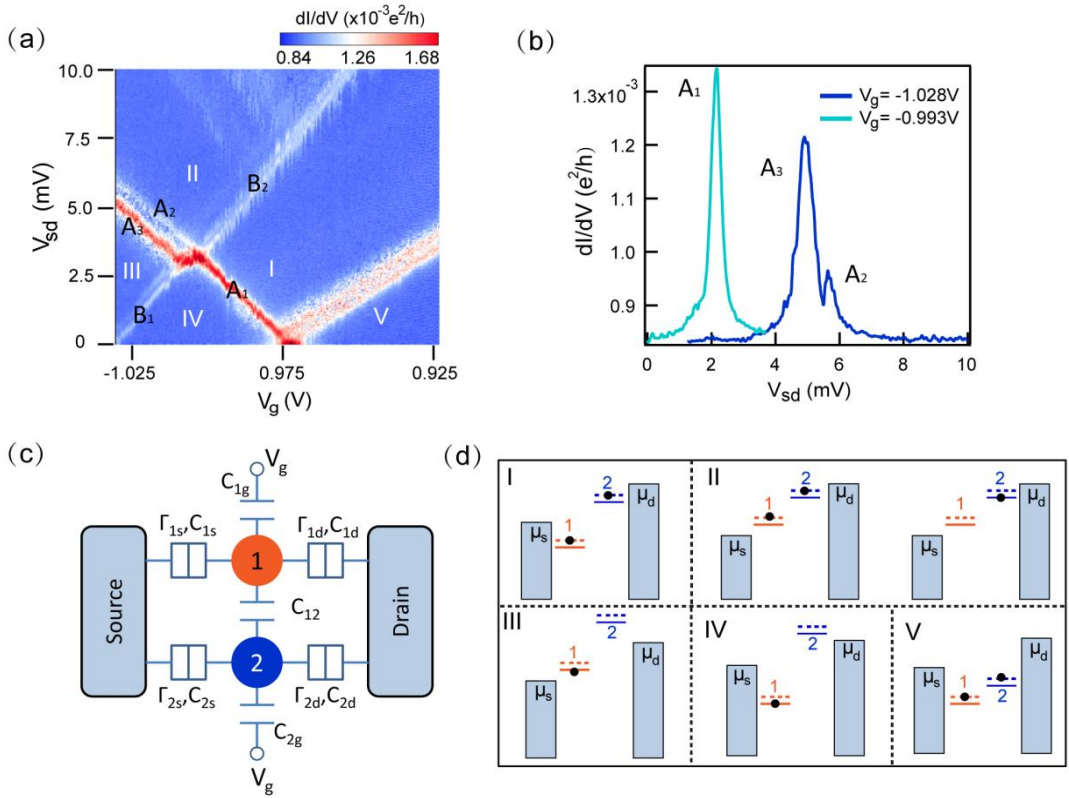

**Figure 2** (a) Detailed measurement for the dashed rectangle region in Figure 1(b). Different regions in the transport map are marked by I,II,III,IV and V, which are associated with different situations of charge transport through the two molecules. II and III (I and II) are the sequential electron tunneling (SET) region of molecule 1(2), in which single electron transport through molecule 1(2) sequentially. In other regions, molecule 1(2) is blocked. $A_1, A_2$ and $A_3$ are conduce peaks of molecule 2 with negative slope. $B_1$ and $B_2$ are conduce peaks of molecule 1 with positive slope. (b) Line traces of dI/dV against $V_{sd}$ in (a) at two gate voltages, revealing the peak height of $A_1$ $A_2$ and $A_3$. (c) Sketch of the parallel double quantum dot model. Electrons tunnel with rates $\Gamma_{1(2)s(d)}$ between molecule 1(2) to source(drain) electrode. Molecule 1(2) couples to source drain and gate electrodes by capacitance $C_{1(2)s}$, $C_{1(2)d}$ and $C_{1(2)g}$ respectively. The mutual capacitance between two molecules is $C_{12}$ (d) The energy diagrams at position I,II,III,IV and V in (a). $\mu_S$ and $\mu_D$ are the Fermi level of source and drain electrodes and they are biased by $V_{sd}$. When molecule 1(2) is charged by one more electron, the level of molecule 2(1) is shifted from solid line to dash line.

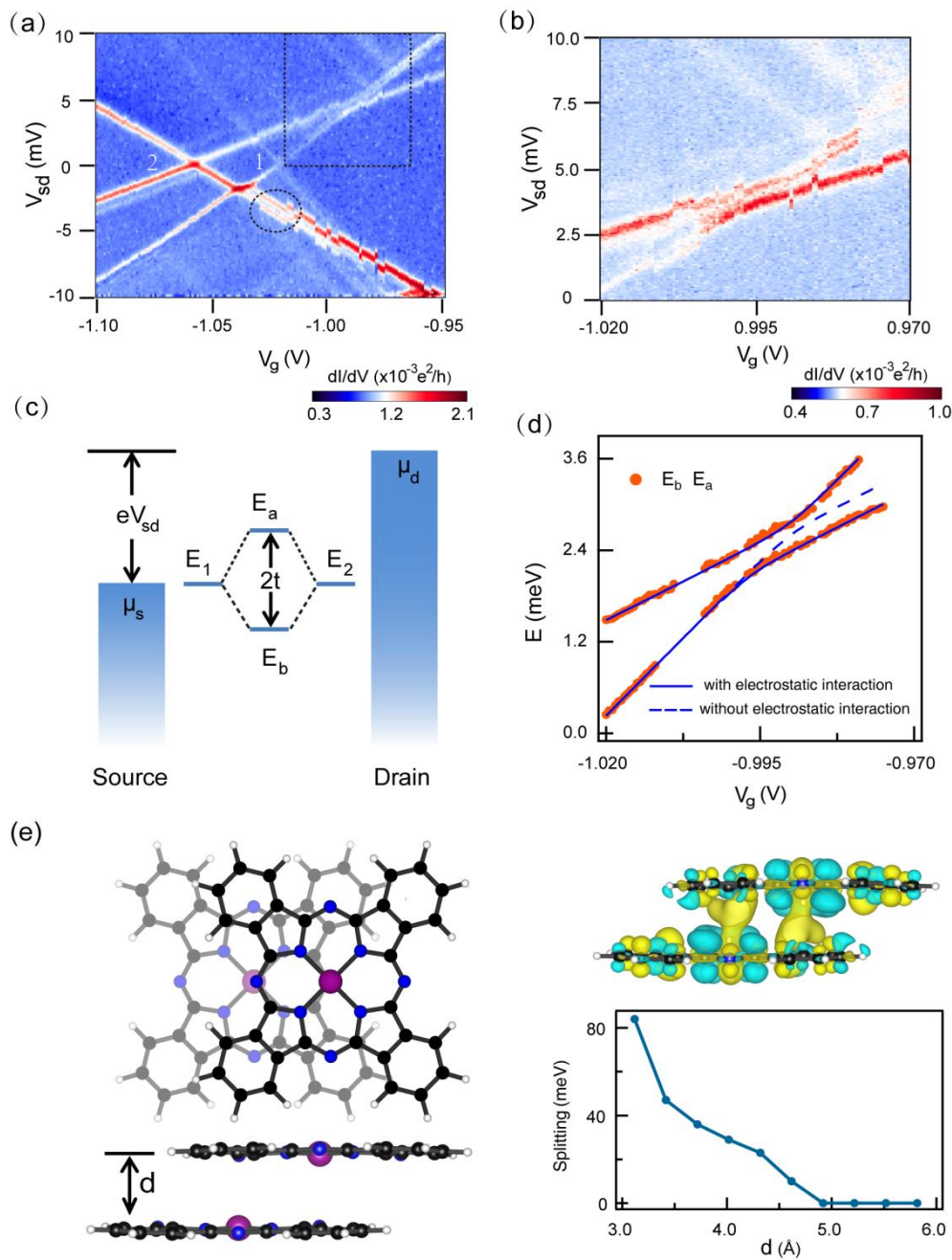

**Figure 3** (a) Color plots of differential conductance (dI/dV) as a function of bias voltage ($V_{sd}$) and gate voltage ($V_g$) for the identical device in Figure1(b) after few times of thermal annealing. The level shift induced by electrostatic interaction is marked by a dashed circle. (b) Close-up of the dashed rectangle region in (a). (c) The energy diagram of the device when two orbitals($E_1$ and $E_2$) are tuned to align with the Fermi surface of source electrode. Two orbitals hybridize with each other and generates a bonding($E_b$) and an antibonding states($E_a$),with an energy separation 2t when$E_1$ and $E_2$ are degenerate. t is the hybridization strength. $\mu_S$ and $\mu_D$ are the Fermi levels of source and drain electrodes, and they are biased by $V_{sd}$ (d) Energy plot of the hybridize orbitals, $E_a$ and $E_b$, close to the degenerate point . They are converted from the positions of the anti-crossing conductance peaks in (b). The peaks in (b) are resonance with source contacts, so the bias voltage can be converted into energy of molecular orbitals by multiplying a factor, ($C_{id}$+$C_{ig}$)/$C_i$ (i=1 or 2 for molecule 1 or 2). Considering the factors for two molecules are very close(0.581 for molecule1 and 0.606 for molecule2), we assume the factors to be same and use the average value(0.594) to simplify calculation. The solid lines are fitting result by Equation1. The dash line indicates energy of $E_b$ without electrostatic interaction. (e) (left panel)One possible of coupled MnPc molecules from top and side views, with the Mn atom in the upper molecule residing on the top the N atom of the lower one. d is the vertical distance between two molecule This configuration is the normal stacking way for MnPc molecules in crystal. (right panel) The calculated charge density difference ($\Delta\rho$= $\rho_{tot}$−$\rho_{upper}$−$\rho_{lower}$) diagrams of hybridized molecules for the configuration in left with d = 3.117 Å. The yellow (blue) region represents acceleration (depletion) electrons and the isosurface level set to be 3.35×10$^{-4}$ e/bohr$^3$. The calculated energy splitting of hybridized orbitals close to Fermi level as a function of distance d.